\documentclass[prl,twocolumn,superscriptaddress,letterpaper, showpacs,amsmath,amssymb,floatfix,nofootinbib]{revtex4-1}
\usepackage{graphicx,bm,url,braket,xcolor,xspace}
\usepackage{hyperref}
\newcommand{\mev}{\, \text{MeV}}
\newcommand{\svec}[1]{\bm{#1}}
\newcommand{\ivec}[1]{\vec{#1}}

\newcommand{\efm}{\,e\,fm$^3$}

\newcommand{\phm}{\phantom{$-$}}

\begin{document}

\title{Correlating Schiff moments in the light actinides with octupole moments}

\author{Jacek Dobaczewski}

\affiliation{Department of Physics, University of York, Heslington, York YO10 5DD, United Kingdom}

\affiliation{Department of Physics, PO Box 35 (YFL), FI-40014 University of
Jyv\"{a}skyl\"{a}, Finland}

\affiliation{Institute of Theoretical Physics, Faculty of Physics, University of Warsaw, ul. Pasteura 5,
PL-02-093 Warsaw, Poland}

\affiliation{Helsinki Institute of Physics, P.O. Box 64, FI-00014 University of
Helsinki, Finland}

\author{Jonathan Engel}

\affiliation{Department of Physics and Astronomy, University of North Carolina,
Chapel Hill, NC, 27516-3255, USA}

\author{Markus Kortelainen}

\affiliation{Department of Physics, PO Box 35 (YFL), FI-40014 University of
Jyv\"{a}skyl\"{a}, Finland}

\affiliation{Helsinki Institute of Physics, P.O. Box 64, FI-00014 University of
Helsinki, Finland}

\author{Pierre Becker}

\affiliation{Department of Physics, University of York, Heslington, York YO10 5DD, United Kingdom}

\email{jacek.dobaczewski@york.ac.uk}

\begin{abstract}
We show that the measured intrinsic octupole moments of $^{220}$Rn, $^{224}$Ra,
and $^{226}$Ra constrain the intrinsic Schiff moments of $^{225}$Ra$^{221}$Rn,
$^{223}$Rn, $^{223}$Fr, $^{225}$Ra, and $^{229}$Pa.  The result is a
dramatically reduced uncertainty in intrinsic Schiff moments.  Direct
measurements of octupole moments in odd nuclei will reduce the uncertainty even
more. The only significant source of nuclear-physics error in the laboratory
Schiff moments will then be the intrinsic matrix elements of the time-reversal
non-invariant interaction produced by CP-violating fundamental physics.  Those
matrix elements are also correlated with octupole moments, but with a larger
systematic uncertainty.
\end{abstract}
 \date{\today}
\maketitle

The observation of a non-zero electric dipole moment (EDM) in a particle, atom,
or molecule with a non-degenerate ground state would signal the violation of
time-reversal (T) symmetry, which in any realistic field theory implies the
violation of charge-parity (CP) symmetry.  The Standard Model violates both
symmetries, of course, but at a level too low to be responsible for the lack of
antimatter in the universe around us \cite{(Mor12),(Din03)}.  The asymmetry
between matter and antimatter is apparently due to a stronger source of CP
violation, and the measurement of an atomic EDM is one of our best hopes to
discover that source.

The atomic isotope with the best limit on its EDM is currently $^{199}$Hg.
About 20 years ago, however, it was realized \cite{(Aue96a),(Eng00a)} that atoms
whose nuclei are asymmetrically shaped (octupole deformed, like pears) would
have enhanced EDMs if the CP violation occurred within the nucleus.  The reason
is connected to a partial screening of nuclear EDMs by electrons
\cite{(Sch63a)}.  The argument goes as follows:

Because of the screening, the nuclear quantity that induces the atomic EDM is
not the nuclear EDM itself, but rather the nuclear Schiff moment:
\begin{equation}
\label{eq:def}
S \equiv \langle \Psi_0 | \hat{S}_0 | \Psi_0 \rangle \approx \sum_{i \neq 0}
         \frac{\langle \Psi_0 | \hat{S}_0 |\Psi_i \rangle
               \langle \Psi_i | \hat{V}_{PT} | \Psi_0 \rangle}
              {E_0 - E_i}
         + \text{c.c.} \,,
\end{equation}
where $| \Psi_0 \rangle$ is the member of the ground-state multiplet with the
maximum angular-momentum $z$-projection, $\ket{\Psi_i}$ are excited states
having the same angular-momentum quantum numbers as the ground state but
opposite parity, and $\hat{S}_0$ is the Schiff operator
\begin{equation}
\label{eq:schiff-op}
\hat{S}_0= \frac{e}{10}\sqrt{\frac{4\pi}{3}}\sum_i  \left( r_i^3
- \frac{5}{3} \overline{r_{\rm ch}^2} r_i \right) Y^1_0(\Omega_i) + \ldots
\,.
\end{equation}
Here the sum is over protons, $\overline{r_{\rm ch}^2}$ is the mean-square
charge radius, and the omitted terms are smaller \cite{senkov08} and, to some
extent, in dispute \cite{liu07,senkov08}. The operator $\hat{V}_{PT}$ in
Eq.~(\ref{eq:def}) is the CP-violating nucleon-nucleon interaction, to be
discussed shortly.  The asymmetric shape of octupole-deformed nuclei implies
parity doubling (see, e.g., Ref.~\cite{(She89a)}): the presence of a partner
$\ket{\overline{\Psi}_0}$ for the ground state $\ket{\Psi_0}$ --- with the same
intrinsic structure and angular momentum but opposite parity --- at a low
excitation energy $\Delta E$. In $^{225}$Ra, for example, the $1/2^+$ ground
state has a $1/2^-$ partner at 55\,keV~\cite{(Hel87a)}.  The similarity of the
two partner states and the low excitation energy means not only that the partner
dominates the sum in Eq.~\eqref{eq:def}, leading to the quite accurate
approximation,
\begin{equation}
\label{eq:S-doublet}
S \approx - 2\frac{\bra{\Psi_0} \hat{S_0}\ket{\overline{\Psi}_0}
 \bra{\overline{\Psi}_0}\hat{V}_{PT} \ket{\Psi_0}}{\Delta E} \,,
\end{equation}
but also that it enhances the Schiff moment by large amounts over the moments in
nuclei with symmetric shapes.

Much of the enhancement is due to the small energy denominator $\Delta E$, but
some comes from the presence in the numerator of the Schiff operator rather than
the electric dipole operator.  Dipole moments are delicate because they depend
on the difference between the center of mass and center of charge, which is
often small even in octupole-deformed nuclei.  Because of the radial weighting
in Eq.~\eqref{eq:schiff-op}, however, Schiff moments can be substantial even if
the centers of mass and charge coincide.

The expectation value of the first term in Eq.~\eqref{eq:schiff-op} is much
larger in octupole-deformed nuclei than that of the second term, which is
proportional to the EDM.  In fact, if the spherical harmonic $Y^1_0$ were
replaced by $Y^3_0$, that first term would just be proportional to the octupole
charge operator%
\footnote{In this work, we use definition of Ref.~\cite{(Boh69)}.  The
definitions ${\hat Q}^3_0 = e\sqrt{4\pi/7} \sum_i r_i^3
Y^3_0(i)$~\protect\cite{(Rob12)} and ${\hat Q}^3_0 = e\sqrt{16\pi/7} \sum_i
r_i^3 Y^3_0(i)$~\protect\cite{(Lea88),(But96)} also appear in the literature.
The experimental value of the $^{224}$Ra spectroscopic octupole moment measured
in Ref~\cite{(Gaf13)}, 2520$\pm$90{\efm} requires the definition of
Refs.~\cite{(Lea88),(But96)}. With our definition, this number becomes
$Q^3_0=940\pm$30{\efm}.}: ${\hat Q}^3_0 \equiv e\sum_i r_i^3 Y^3_0(i)$, where
the sum again is over protons.  The matrix elements of this operator are a
direct measure of octupole deformation and thus not at all delicate in
octupole-shaped isotopes.

This argument has an obvious implication: the Schiff moment should be correlated
with the matrix elements of the octupole moment, and measured octupole
transition rates should allow us to reduce the uncertainty in calculations of
Schiff moments.  Such calculations are essential if we want to use limits on
atomic EDMs (or an eventual observation of one) to make quantitative statements
about new sources of CP violation.  Existing calculations \cite{(Dob05g)},
carry an uncertainty \cite{(Eng13)} that is significantly larger than $100\%$,
and the use of complementary measurements to exploit them is important.  In the
rest of this paper, we show that measured octupole properties are a great help.

The story is not quite as simple as it first appears, however, partly because
the Schiff operator is not the only ingredient in Eq.~\eqref{eq:def}; the
CP-violating potential $V_{PT}$, which atomic EDM experiments hope to elucidate,
also plays a role. The potential, which is often discussed in terms of meson
exchange, can be represented in chiral effective field theory \cite{Maekawa2011},
a QCD-based picture of interacting nucleons and pions that has a systematic
power-counting scheme.  Including the most important terms, one has
\cite{(Hax83a),(Her88a),Maekawa2011},
\begin{align}
\label{eq:pion}
&\hat{V}_{PT}(\svec{r}_1-\svec{r}_2)
                = -   \frac{g \, m_{\pi}^2}{8 \pi m_N}
                     \Big\{
                         (\svec{\sigma}_1 - \svec{\sigma}_2) \cdot
                         (\svec{r}_1-\svec{r}_2) \\
                         &\times \left[ \bar{g}_0 \, \ivec{\tau}_1 \cdot
                                             \ivec{\tau}_2
                         - \frac{\bar{g}_1}{2} \,
                          (\tau_{1z}+\tau_{2z}) + \bar{g}_2
                         (3\tau_{1z}\tau_{2z} - \ivec{\tau}_1 \cdot
                         \ivec{\tau}_2 ) \right]  \nonumber \\
                         &\quad \quad -\frac{\bar{g}_1}{2}
                         (\svec{\sigma}_1+\svec{\sigma}_2)
                         \cdot (\svec{r}_1-\svec{r}_2) \,
                         (\tau_{1z}-\tau_{2z})  \Big\} \nonumber \\
& \times \frac{{\rm exp}(-m_{\pi} |\svec{r}_1-\svec{r}_2|)}{m_{\pi}
|\svec{r}_1-\svec{r}_2|^2} \left[ 1+\frac{1}{m_{\pi}|\svec{r}_1-\svec{r}_2|}
\right]  \nonumber \\
&+ \frac{1}{2m_N^3}\left[ \bar{c}_1 + \bar{c}_2 \ivec{\tau}_1 \cdot
\ivec{\tau}_2  \right] \left( \svec{\sigma}_1 - \svec{\sigma}_2 \right) \cdot
\bm\nabla \delta^3(\svec{r}_1 -\svec{r}_2) \,, \nonumber
\end{align}
where arrows denote isovector operators, $\tau_z$ is +1 for neutrons, $m_N$ is
the nucleon mass, and (in this equation only) we use the convention \mbox{$\hbar
= c = 1$}.  The $\bar{g}$'s are the unknown isoscalar, isovector, and isotensor
$T$-violating pion-nucleon coupling constants, the $\bar{c}$'s are the unknown
coupling constants of a short-range interaction that subsumes the effects of
heavy-meson exchange, and $g$ is the usual strong ${\pi NN}$ coupling constant.
Most calculations thus far have neglected the effects of the contact
interactions in the last line of Eq.~\eqref{eq:pion} (as well as the effects of
neutron and proton EDMs). Here we write the Schiff moment $S$ in
Eq.~\eqref{eq:def} as
\begin{equation}
\label{eq:coefs}
S = a_0 \, g \, \bar{g}_0 + a_1 \, g \,  \bar{g}_1 +a_2 \, g \,  \bar{g}_2
  + b_1 \,      \bar{c}_1  +b_2 \,       \bar{c}_2 .
\end{equation}
The coefficients $a_i$ and $b_i$, which are the result of a calculation, have
units\,{\efm}.

Another slight complication comes from our use of Skyrme \cite{skyrme58,(Ben03e)}
or Gogny \cite{decharge80,(Ben03e)} energy-density functional theory (related to
mean-field theory) to express the nuclear wave function in terms of a deformed
and parity-mixed Slater determinant or a more general deformed quasiparticle
vacuum.  The deformed wave function represents the \textit{intrinsic} state of
the nucleus $\ket{\Phi_0}$, that is, the nuclear state in a body-fixed frame.
The Schiff moment in this frame is independent of $\hat{V}_{PT}$, the function
of which, in a manner of speaking, is to ensure that the intrinsic breaking of
parity and time-reversal symmetries by mean-field theory survives in the
laboratory frame.  It is the intrinsic Schiff moment that we can most easily
correlate with measured octupole transition rates, as we now explain:

Having obtained an intrinsic state through mean-field-like calculations, one
needs to project it onto laboratory states with well-defined angular momentum
and parity, in our case the two states in the parity doublet that determine the
laboratory Schiff moment.  In our prior work on the subject, described in Ref.\
\cite{(Dob05g)}, and in the measurements of deformation, shapes are assumed to
be infinitely rigid.  This approximation leads to the well-known result
\cite{Boh75} that all of the ground-band reduced matrix elements of an
(arbitrary) operator $\hat{X}^\lambda$ with multipolarity $\lambda$ are
proportional to an intrinsic-state expectation value $\langle \hat{X}^\lambda_0
\rangle$:
\begin{align}
\label{eq:rot-approx}
\bra{J}\!| \hat{X}^\lambda |\!\ket{J'}_{\rm rigid} &= \sqrt{2J'+1} \braket{J'K,
\lambda 0 | JK}
\braket{\hat{X}^\lambda_0} \,,
\end{align}
where $\braket{J'K, \lambda 0 | JK}$ is a Clebsch-Gordan coefficient and the
intrinsic-state expectation value $\braket{X^\lambda_0}$ is evaluated in an
axially-symmetric state having angular-momentum projection $K$ on the symmetry
axis.
The Wigner-Eckart theorem then implies that the observable laboratory transition
matrix elements
\begin{align}
\label{eq:WE}
\bra{JM} \hat{X}^\lambda_\mu \ket{J'M'} &=  \braket{J'M', \lambda\mu | JM}
\frac{ \bra{J}\!| \hat{X}^\lambda |\ket{J'} }{\sqrt{2J+1}},
\end{align}
can be used to extract values of $\langle \hat{X}^\lambda_0 \rangle$ simply in
the rigid-deformation limit.  With the Clebsch-Gordan coefficients evaluated,
Eqs.~(\ref{eq:rot-approx}) and (\ref{eq:WE}) relate
$\bra{\Psi_0}\hat{S}_0\ket{\overline{\Psi}_0}$ to the intrinsic expectation
value $S_0 \equiv \braket{\hat{S_0}}$ and
$\bra{\Psi_0}{\hat{V}}_{PT}\ket{\overline{\Psi}_0}$ to the intrinsic expectation
value $\braket{\hat{V}_{PT}}$:
\begin{align}
\label{eq:intrinsic-stuff1}
\bra{\Psi_0}\hat{S_0}\ket{\overline{\Psi}_0}_{\rm rigid} &= \tfrac{J}{J+1} S_0 , \\
\label{eq:intrinsic-stuff2}
\bra{\overline{\Psi}_0}\hat{V}_{PT}\ket{\Psi_0}_{\rm rigid} &= \braket{\hat{V}_{PT}}
\,,
\end{align}
where Eq.~\eqref{eq:intrinsic-stuff1} is specified for $J=J'=M=M'=K$.
Eqs.~\eqref{eq:rot-approx} and \eqref{eq:WE} also relate octupole transition
rates to the intrinsic octupole moment $Q^3_0 \equiv \braket{\hat{Q}^3_0}$.

In fact, we are no longer confined to this rigid-deformation limit; we can now
obtain the ground state and its partner by exactly projecting the lowest
intrinsic state onto states with any angular momentum and with positive or
negative parity.  Thus, we can test the quality of the rigid-deformation
approximation by comparing reduced matrix elements from
Eq.~(\ref{eq:rot-approx}) to those evaluated exactly between angular-momentum-
and parity-projected states.  For the Schiff operator the rigid-deformation
approximation turns out to be very good; it induces an error of only about
1.5\%.  The rigid-deformation approximation for the octupole operator
$\hat{Q}^3$ in $^{224}$Ra is even better, inducing an error of less than 0.1\%.

The error from extrapolating our results to a single-particle space with an
infinite number of harmonic-oscillator shells is also quite small, about 0.02\%.
For calculations with 20 oscillator shells, which lead to a reasonable balance
between CPU time and precision, the two errors have similar magnitude and
opposite sign.  We will thus use this basis for all calculations performed with
Skyrme functionals.  The Gogny functional leads to CPU time that is much longer;
we therefore settle for 16 oscillator shells when working with it.  We use the
code {\sc hfodd} (v2.84h)~\cite{(Sch17a),(Dob18)} to carry out all Hartree-Fock
(HF), Hartree-Fock-Bogolyubov (HFB), or BCS calculations.  The accuracy of all
these approximations means that we can consider the Schiff and octupole
transition matrix elements to be directly proportional to the corresponding
intrinsic moments, the correlation of which we now address in more detail.

\begin{figure}[t]
\centering
\includegraphics[width=\columnwidth]{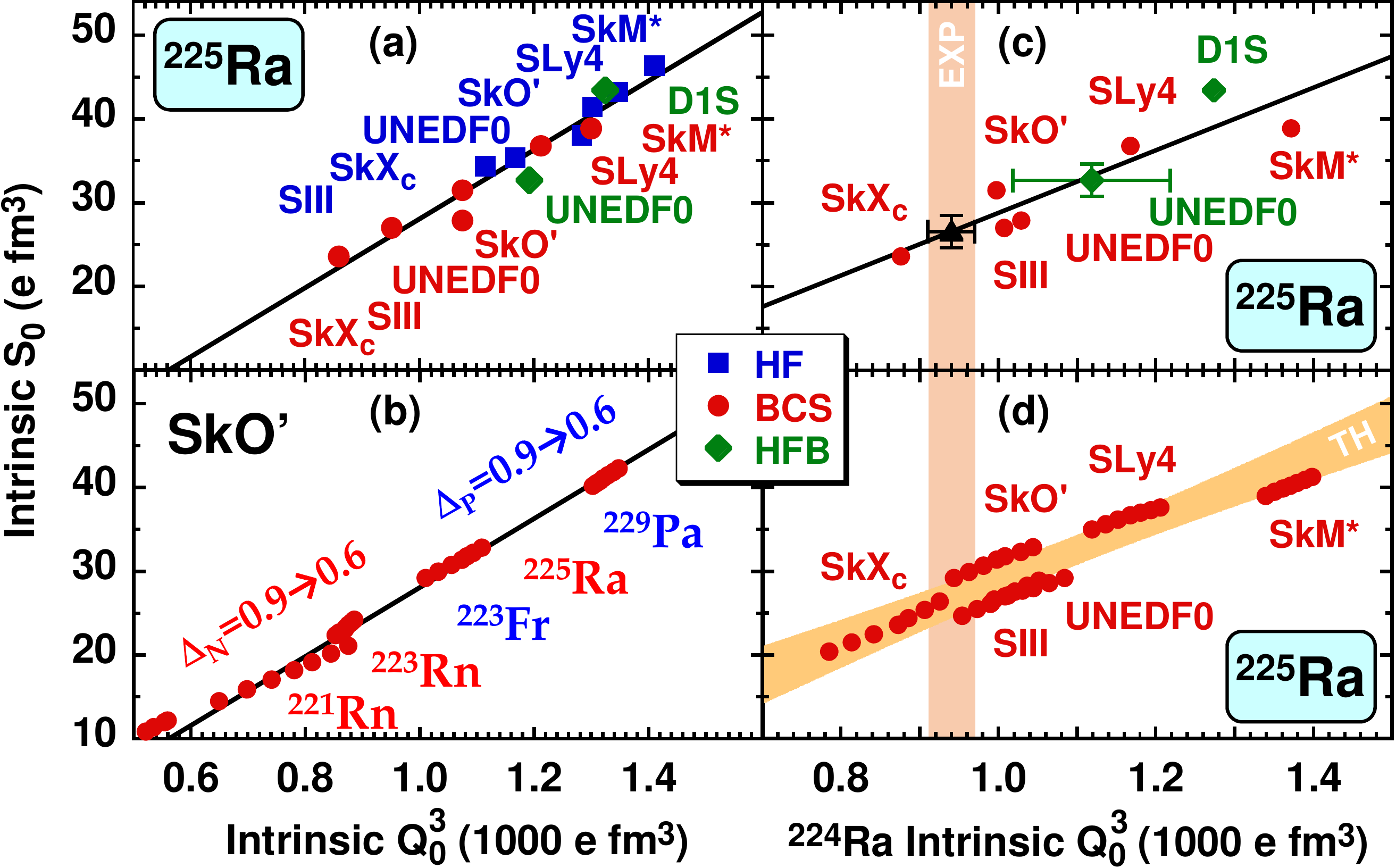}
\caption{\label{fig:schiff-many} (a) The unconstrained predictions of the
functionals SkX$_c$~\cite{(Bro98)}, SIII~\cite{(Bei75b)}, UNEDF0~\cite{(Kor10c)},
SkO$^\prime$~\cite{(Rei99)}, SLy4~\cite{(Cha98a)}, SkM*~\cite{(Bar82c)}, and
D1S~\cite{(Ber91c)} for the octupole and Schiff moments of $^{225}$Ra, with and
without pairing. (b) Predictions of SkO$^\prime$ for the Schiff and octupole
moments in $^{221}$Rn, $^{223}$Rn, $^{223}$Fr, $^{225}$Ra, and $^{229}$Pa, with
varying amounts of neutron ($N$) or proton ($P$) pairing. In odd-$N$ or odd-$Z$
nuclei, the moments increase with as the pairing gaps decrease from
$\Delta_{N(P)}=0.90$ to $0.70$\,MeV, in steps of 0.05\,MeV.  Panels (c) and (d)
are like (a) and (b) but with the octupole moment of $^{224}$Ra on the abscissa.
The error bars represent the result of a regression
analysis~\protect\cite{suppl-octuschiff}.}
\end{figure}

Figure \ref{fig:schiff-many}(a) shows the unconstrained predictions of many
functionals, with and without pairing (that is, in the HF or HFB/BCS
approximations), for the octupole and Schiff moments of $^{225}$Ra.  The
correlation between the two observables is striking, good enough so that one can
nearly identify a unique prediction for the intrinsic Schiff moment, given a
measured octupole moment.  We say ``nearly'' because neutron pairing, the exact
strength of which is unknown, introduces some ambiguity.  The red dots
correspond to a pairing gap of $0.747$ MeV, a sensible guess.
Figure \ref{fig:schiff-many}(b) shows the range of predictions with the
functional SkO$^\prime$ for the reasonable range $0.6 \mev \le \Delta_{N(P)} \le
0.9 \mev$ in five odd-$A$ nuclei with large asymmetric deformation.  The correlation
between Schiff and octupole moments is quite linear and it allows the
identification of an uncertainty for the predictions SkO$^\prime$.

All these results indicate the desirability of measuring $Q^3_0$ in odd nuclei
for which atomic EDM measurements are conceivable.  Though no one has made such
a measurement (yet), the Liverpool group reported the measurement of $Q^3_0$ in
the neighboring even-even nucleus $^{224}$Ra a few years ago \cite{(Gaf13)}.
Figures~\ref{fig:schiff-many}(c) and (d) show the same kind of results as do
Figs.~\ref{fig:schiff-many}(a) and (b), but with the predicted intrinsic Schiff
moments plotted versus the octupole moments of $^{224}$Ra. The vertical band in
the figures represents the measured value of the $^{224}$Ra octupole moment,
with the width of the bar representing experimental uncertainty. The tilted line
in Fig.~\ref{fig:schiff-many}(c) and band in Fig.~\ref{fig:schiff-many}(d)
represent, respectively, the correlation and total
uncertainty~\cite{suppl-octuschiff}\nocite{(Dob14b),(Bir32a),(She95)} of the Schiff moment.
Figure~\ref{fig:schiff-many}(d) also shows that the correlation holds not only
for different functionals, but also for varying pairing strengths.

Figure~\ref{fig:schiff-many}(c) also shows the results of a quantitative
analysis.  We use linear regression to determine the coefficients $a$ and $b$ in
the relation $S_{10}=a+b \times Q^3_0(^{224}\text{Ra})$.  The Supplemental
Material~\cite{suppl-octuschiff} for this manuscript
provides more details. For $^{225}$Ra, the propagated intrinsic Schiff moment
and its uncertainty at the experimental intrinsic octupole moment
$Q^3_0(^{224}\text{Ra})=940(30)${\efm} is $S_{0}=26.6(1.9)${\efm}. The
theoretical uncertainty of 1.6{\efm} is larger than that from experiment, which
is 1.1{\efm}.

\begin{figure}[t]
\centering
\includegraphics[width=\columnwidth]{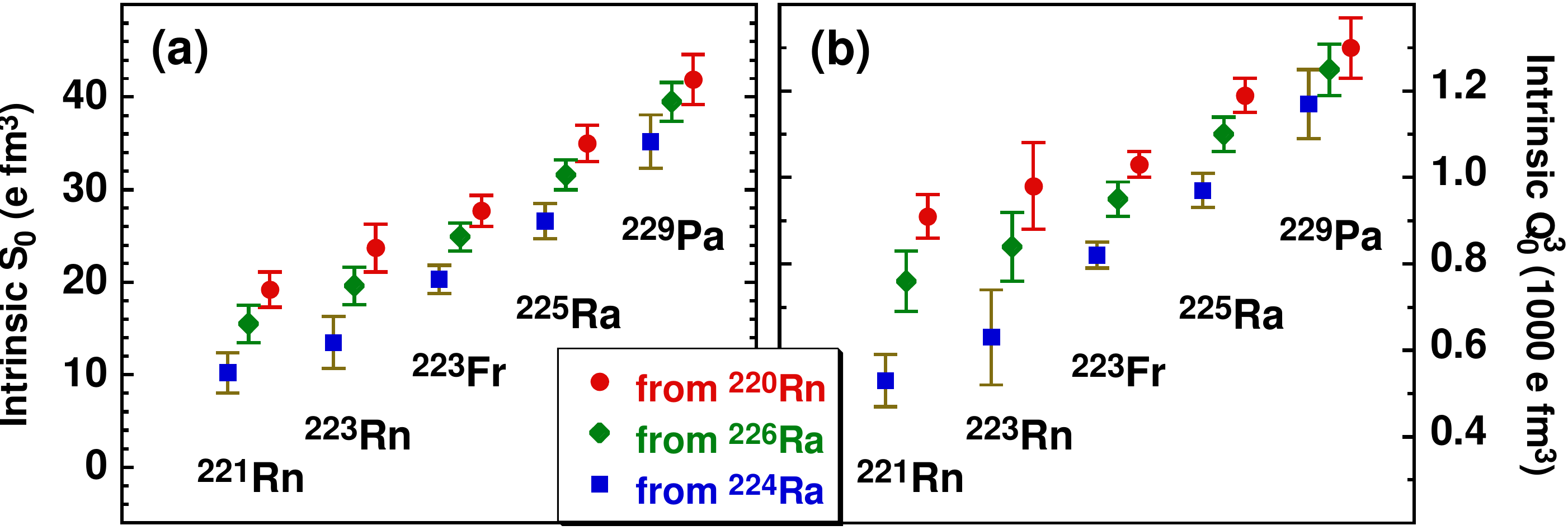}
\caption{\label{fig:final}(Color online) Intrinsic Schiff moments $S_{0}$ in
{\efm} (a) and octupole moments $Q^3_0$ in units of 1000\,{\efm} (b) of
$^{221}$Rn, $^{223}$Rn, $^{223}$Fr, $^{225}$Ra, and $^{229}$Pa, determined from
the experimental octupole moments of $^{224}$Ra, $^{226}$Ra, and $^{220}$Rn.  }
\end{figure}

It is now clear that the observed correlation between the calculated intrinsic
Schiff moment in $^{225}$Ra and octupole moment in $^{224}$Ra allows us to
greatly reduce systematic uncertainties stemming from nuclear functionals.
Figure~\ref{fig:final}(a) shows predictions for the intrinsic Schiff moments of
$^{221}$Rn, $^{223}$Rn, $^{223}$Fr, $^{225}$Ra, and $^{229}$Pa from the
experimental octupole moments of $^{224}$Ra~\cite{(Gaf13)},
$^{226}$Ra~\cite{(Wol93)}, and $^{220}$Rn~\cite{(Gaf13)}. A similar analysis,
shown in Fig.~\ref{fig:final}(b), allows us to predict values of octupole
moments in these odd nuclei.  Numerical values for all these intrinsic moments
are collected in the Supplemental Material~\cite{suppl-octuschiff}.

\begin{figure}[b]
\centering
\includegraphics[width=\columnwidth]{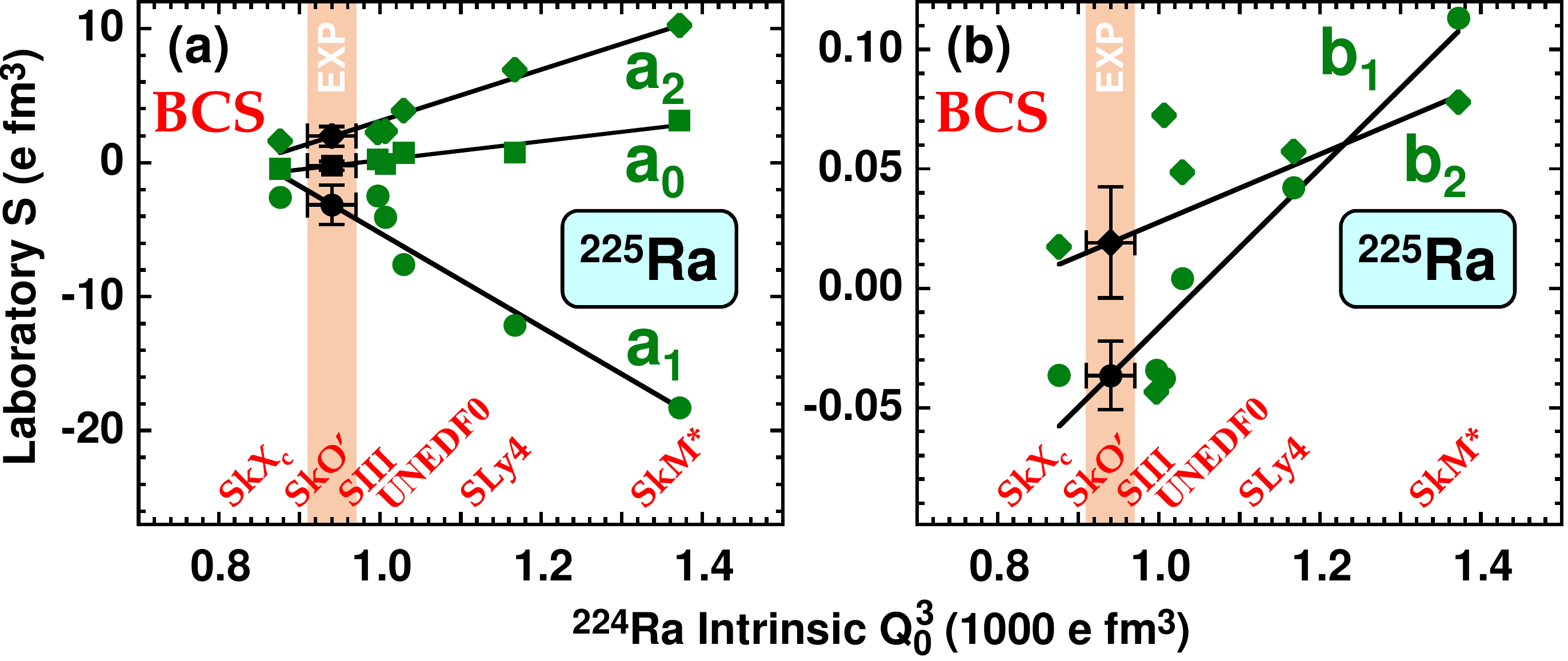}
\caption{\label{fig:g-var}(Color online) Coefficients $a_0$, $a_1$, and $a_2$
(a) and $b_1$ and $b_2$ (b), Eq.~(\protect\ref{eq:coefs}), corresponding to the
finite and zero-range terms of ${\hat{V}}_{PT}$, Eq.~(\protect\ref{eq:pion}),
determined in $^{225}$Ra for six Skyrme functionals and propagated to the
experimental value of the octupole moment in $^{224}$Ra.}
\end{figure}

To obtain an independent estimate of systematic uncertainties in the intrinsic
Schiff moment of $^{225}$Ra, we employ the full covariance matrix of the UNEDF0 functional
model parameters~\cite{(Kor10c)}. This gives an intrinsic Schiff moment in
$^{225}$Ra of $S_{10}=32.7(1.9)${\efm} and an octupole moment in $^{224}$Ra of
$Q^3_0=1.17(10) \times \,1000$\,{\efm} (cf.\ the error bars in
Fig.~\ref{fig:schiff-many}) (c)). It also yields a very strong correlation coefficient of 0.908 between
the two moments. The relatively large uncertainty in
$Q^3_0$ means that only a modest increase in the UNEDF0 penalty function is
required to alter the parameters of the coupling constants so that the
calculated $Q^3_0$ agrees with experiment.  The strong correlation between the
octupole and Schiff moments then means that the Schiff moment would probably
slide closer to our propagated value of 26.6{\efm}. This hypothetical result,
however, can only be verified by a full refit of UNEDF0 with the experimental
value of $Q^3_0$ in $^{224}$Ra included in the penalty function.

So far all our focus has been on the intrinsic Schiff moment, which, as we have
noted, is only one of the two ingredients in the laboratory Schiff moment
(\ref{eq:S-doublet}); the other is the intrinsic matrix element of
${\hat{V}}_{PT}$. Can we use measured octupole moments to constrain it as well?
Fig.~\ref{fig:g-var} shows the variation of coefficients $a_0$, $a_1$, $a_2$,
$b_1$, and $b_2$, Eq.~(\protect\ref{eq:coefs}),\footnote{In
Refs.~\cite{(Eng03b),(Dob05g)}, signs of coefficients $a_0$, $a_1$, and $a_2$
were inverted.} in $^{225}$Ra with the octupole moment in $^{224}$Ra.
Apart from $b_2$, there is a clear correlation that allows for a
meaningful extrapolation to the measured value; a larger scatter of points
induces a larger extrapolation error of $b_2$.

The analysis becomes more complicated, however, when we include measured
octupole moments from other isotopes.  The correlation between a given octupole
moment and the coefficients $a_i, b_i$ still exists, but the use of two
different octupole moments to constrain the coefficients can lead to quite
different values.  This situation is unlike that depicted by Fig.\
\ref{fig:final}, and suggests the presence of significant systematic error in
the calculations of the intrinsic matrix element of $\hat{V}_{PT}$.

The figures in our Supplemental Material~\cite{suppl-octuschiff} show that the
correlation between an octupole moment in one nucleus and a laboratory Schiff
moment in another is better if the two nuclei are very close together in $Z$ and
$N$.  We therefore use only the octupole moment of $^{220}$Ra in computing the
coefficients $a_i$ and $b_i$ in $^{221}$Rn and $^{223}$Rn, and only the moments
of $^{224}$Ra and $^{226}$Ra when computing the coefficients in $^{223}$Fr,
$^{225}$Ra, and $^{229}$Pa.  As Table~\ref{tab:final} shows, we still end up
with a sizable uncertainty in the coefficients, even with these restrictions.
The numbers in (blue) italics there are consistent with zero, and only those in
(red) boldface are determined with a precision of 25\% or better.  In
$^{225}$Ra, our central values for $a_1$ and $a_2$ are slightly smaller than in
our earlier computation \cite{(Dob05g)}, while $a_0$ is consistent with zero.
Note that only in this nucleus and in $^{223}$Fr, were we able to use the
experimental excitation energy in Eq.~\eqref{eq:S-doublet}; for the other
isotopes, because reliable measurements are not available, we took the
excitation energy to be 100\,keV.  As Eq.~(\ref{eq:S-doublet}) shows, when data
become available, the results in the table can be simply scaled.

\begin{table}[t]
\caption{(Color online) Coefficients $a_0$, $a_1$, $a_2$, $b_1$, and $b_2$ (in
{\efm}) (from Eq.~(\protect\ref{eq:coefs})), determined by regression analysis.
For $^{221}$Rn and $^{223}$Rn we show values propagated to the experimental
octupole moment of $^{220}$Rn, whereas for $^{223}$Fr, $^{225}$Ra, and
$^{229}$Pa we show averages of those propagated to $^{224}$Ra and $^{226}$Ra.
Details are in the Supplemental Material~\protect\cite{suppl-octuschiff}. Values
determined with a precision better than 25\% are in (red) boldface and those
compatible with zero are in (blue) italics.}
\label{tab:final}
\begin{tabular}{@{}l|l@{}l@{}l@{}l@{}l@{}}
\hline
  & \multicolumn{1}{c}{$a_0$} & \multicolumn{1}{c}{$a_1$} &
  \multicolumn{1}{c}{$a_2$} & \multicolumn{1}{c}{$b_1$} &
  \multicolumn{1}{c}{$b_2$}  \\
\hline
$^{221}$Rn &   $-${\color {blue}\it0.04(10) } &   $-${\color  {red}\bf1.7(3) } & \phm{\color  {red}\bf 0.67(10) } &      $-${0.015(5) } &      $-${0.007(4)}       \\
$^{223}$Rn &   $-${\color {blue}\it0.08(8)  } &   $-${\color  {red}\bf2.4(4) } & \phm{\color  {red}\bf 0.86(10) } &      $-${0.031(9) } &   $-${\color {blue}\it0.008(8)}       \\
$^{223}$Fr &  \phm{\color {blue}\it0.07(20) } &   $-${\color{ blue}\it0.8(7) } & \phm{\color {blue}\it 0.05(40) } &     \phm{0.018(8) } &      $-${0.016(10)}      \\
$^{225}$Ra &  \phm{\color {blue}\it0.2(6)   } &      $-${5(3)   } &     \phm{3.3(1.5) } &   $-${\color {blue}\it0.01(3)  } &     \phm{0.03(2)}        \\
$^{229}$Pa &   $-${\color  {red}\bf1.2(3)   } &  \phm{\color {blue}\it0.9(9) } &  $-${\color {blue}\it 0.3(5)   } &  \phm{\color  {red}\bf0.036(8) } &     \phm{0.032(18)}      \\
\hline
\end{tabular}
\end{table}

How does one reduce the uncertainty in the laboratory Schiff moments?  The
obvious way is to isolate and measure a quantity that is closely correlated with
the intrinsic matrix element of $\hat{V}_{PT}$. Although that potential is a
two-body operator, it can be approximated by an average one-body operator with
the schematic form $\svec{\sigma} \cdot \svec{r}$, as, e.g., in Refs.\
\cite{(Aue96a)} and \cite{(Eng03b)}.  Such operators (and related two-body
meson-exchange versions) occur within subleading pieces of the hadronic
electro-weak current, but identifying and measuring the appropriate matrix
elements will be a challenge.  The potential payoff, however, makes it worth
addressing.

We thank Michael Ramsey-Musolf and Peter Butler for useful discussions.  This
work was supported by the U.S.\ Department of Energy through Contract No.\
DE-FG02-97ER41019, by the STFC grants Nos.~ST/M006433/1 and ST/P003885/1, and by
the Academy of Finland and University of Jyv\"askyl\"a within the FIDIPRO
program.  We acknowledge the CSC-IT Center for Science Ltd., Finland, for the
allocation of computational resources.


%

\end{document}